\documentclass[twocolumn, prl,showpacs]{revtex4}
\usepackage{graphicx}

\begin{document}

\title{Controlling high-frequency collective electron dynamics via single-particle complexity} 

\author{N. Alexeeva$^1$,  M.T. Greenaway$^1$, A.G. Balanov$^{2}$, O. Makarovsky$^1$,  A. Patan\`{e}$^1$, M.B. Gaifullin$^2$, F. Kusmartsev$^{2}$ and T.M. Fromhold$^1$}
\affiliation{$^1$School of Physics and Astronomy, University of Nottingham, Nottingham NG7 2RD, United Kingdom \\
$^2$Department of Physics, Loughborough University,
Leicestershire, LE11 3TU, United Kingdom}
\date{\today}


\begin{abstract}
We demonstrate, through experiment and theory, enhanced high-frequency current oscillations due to magnetically-induced conduction resonances in superlattices. Strong increase in the ac power originates from complex single-electron dynamics, characterized by abrupt resonant transitions between unbound and localized trajectories, which trigger and shape propagating charge domains. Our data demonstrate that external fields can tune the collective behavior of quantum particles by imprinting configurable patterns in the single-particle classical phase space. 
\end{abstract}


\maketitle{}
Understanding the interplay between the properties of individual objects and their collective behavior is of fundamental interest in many fields \cite{Aranson,Hunt,Biroli07,Edling11}. It explains, for example, jamming and pattern formation in granular systems \cite{Aranson,Rivas,Clerc,Swift}, dynamical heterogeneity in phase transitions \cite{Biroli07}, tunneling dynamics and quantum phases in cold atoms \cite{Wimberger,Altman}, and the synchronization of networks and complex adaptive systems \cite{Hunt,Zhao}. Moreover, interactions between particles play a key role in determining the structures formed when the particles come into contact \cite{Edling11}. Consequently, tailoring single-particle dynamics may provide a route to controlling the collective dynamics of many-body systems. This is a major challenge both in fundamental science \cite{Hanggi09,Whitelam10,Weit2011} and for developing new technologies such as high-frequency electronic devices \cite{Moss,Ozyuzer,Hyart}, whose performance can be greatly enhanced by applied quantizing magnetic fields \cite{Wade,Scalari}.


The phase space structure of individual particles, in particular the existence and relative location of regular and chaotic trajectories, critically affects thermalization and diffusion both in classical and quantum systems \cite{Zasl91,Cass09}. Therefore, manipulating the single-particle phase space, by generating new chaotic trajectories for example, is a promising strategy in the search for ways to control other collective phenomena.

\begin{figure}
\centering
 \includegraphics*[width=.8\linewidth]{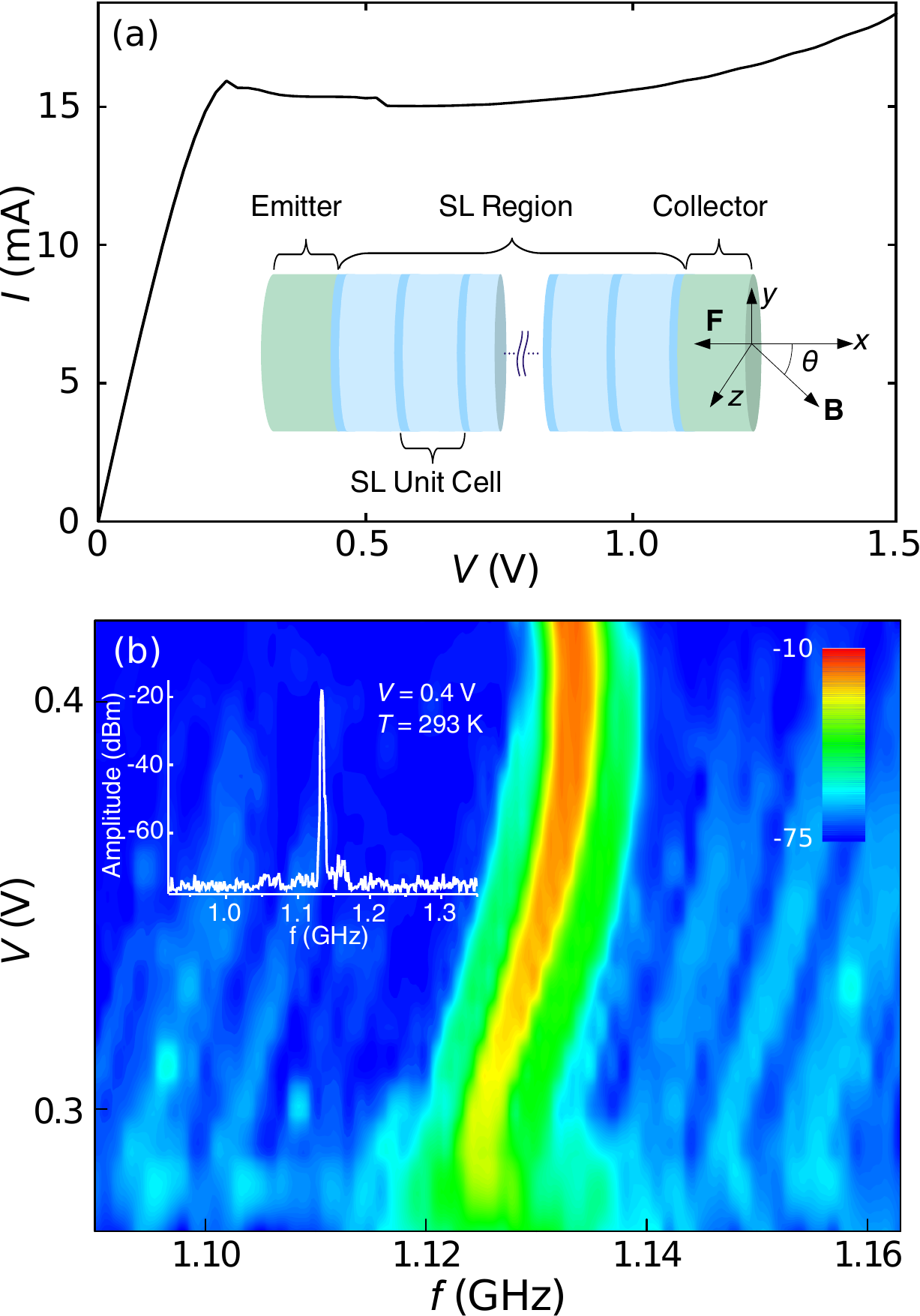}
 \caption{\label{fig:1} (Color) (a) dc $I(V)$ curve measured for the SL when $B$ = 0 at temperature $T = 293$ K. Inset: schematic diagram of the SL showing the emitter/collector contacts (green), barrier and quantum well layers (dark and light blue respectively), and orientation of the applied electric field, $\textbf{F}$, and magnetic field, $\textbf{B}$, relative to the co-ordinate axes. (b) Color map (scale inset right calibrated in dBm) showing the spectrum of $I(t)$ oscillations measured over a range of $V$ and frequencies, $f$, near the first harmonic. Left inset: a typical frequency spectrum of the $I(t)$ oscillations measured around the first harmonic for $V=0.4$ V and $T = 293$ K.}
\end{figure}

Usually, the transition to chaos in Hamiltonian systems occurs by the gradual destruction of stable orbits, in accordance with the Kolmogorov-Arnold-Moser (KAM) theorem \cite{Lichtenberg 1992}. In far rarer non-KAM chaos, the chaotic orbits become abruptly unbounded when the perturbation frequency attains critical values and map out intricate ``stochastic webs'' in phase space \cite{Zasl91}. Experimental realization of non-KAM classical chaos was recently achieved using a quantum system \cite{FRO2004}: a semiconductor superlattice (SL) with a magnetic field, $\textbf{B}$, tilted relative to an electric field, $\textbf{F}$, along the SL axis. When the frequency of single-electron Bloch oscillations along the SL axis is commensurate with that for cyclotron motion in the plane of the layers \cite{FRO2004, Kosevich, Hyart}, the orbits map out stochastic webs in phase space, which delocalize the electrons in real space. This delocalization creates multiple resonant peaks in the single-electron drift velocity versus $F$ characteristics, which enhance the measured dc conductivity \cite{FRO2004}.

In this Letter, we show, via experiments and theoretical modeling, that non-KAM single-particle trajectories have a dramatic effect on the ac collective dynamics of electrons in a SL, thus providing a mechanism for controlling and enhancing the high-frequency (GHz) ac response of this solid-state device. Our work therefore overcomes a major long-standing limitation of SL oscillators, namely how to boost the ac power produced by an \emph{individual} device. At critical applied fields, the single-electron phase space reconfigures into extended stochastic webs, which resonantly increase the electron drift velocity. This boosts the strength and speed of propagating charge domains and, hence, the resulting current oscillations. Such tailoring of the collective electron dynamics is a non-trivial nonlinear effect, which we explain by relating islands of high-power current oscillations directly to the emergence of web patterns in phase space. 

More generally, our results demonstrate that collective effects in periodic quantum systems can be controlled by imprinting specific patterns in the underlying classical phase space. Such control may be realized in a wide range of periodic systems, including quantum cascade lasers \cite{Wade,Scalari} and Bose-Einstein condensates in an optical lattice \cite{Altman,Morsch}, or by applying different perturbations, for example an ac bias voltage in the case of the SL \cite{HYA2009}. 

Our SL was grown by molecular beam epitaxy on a (100)-oriented n-doped GaAs substrate. It comprises 15 unit cells [light and dark blue layers in Fig. \ref{fig:1}(a) inset], which are separated from two heavily n-doped GaAs contacts [green in Fig. \ref{fig:1}(a) inset] by Si-doped GaAs layers of width 50 nm and doping density $1\times10^{17}$ cm$^{-3}$. Each unit cell, of width $d$ and Si doped at $3\times10^{16}$ cm$^{-3}$, is formed by a 1 nm thick AlAs barrier, a 7 nm wide GaAs quantum well (QW) and 0.8 InAs monolayers at the center of each QW. The InAs layer facilitates the direct injection of electrons into the lowest energy miniband, of width $\Delta= 19$ meV, and also creates a large minigap ($\Delta_g = 198$ meV), which prevents interminiband tunneling (see Supplemental Material \cite{Supp} and Ref. \cite{PAT2002}). For electrical measurements, the SL was processed into circular mesa structures of diameter 20 $\mu$m with ohmic contacts to the substrate and top cap layer. Magnetic field studies were performed in a superconductor solenoid magnet with $\textbf{B}$ tilted at an angle $\theta$ to the SL ($x$) axis [Fig. \ref{fig:1}(a) inset]. An Anritsu Spectrum Analyzer (MS2667C) was used to measure the frequency, $f$, and amplitude of GHz current oscillations in the SL.

Figure \ref{fig:1}(a) shows the room temperature ($T=293$ K) current-voltage, $I(V)$, characteristics of the SL measured for $B=0$. The $I(V)$ curve exhibits an ohmic region at low $V$, followed by a current peak and a region of negative differential conductance (NDC) that extends over a wide range of applied bias. The NDC creates regions of high electron density (propagating charge domains) \cite{WAC2002}, which generate self-sustained oscillations \cite{SCHO1997, EIS2010} in $I$ versus time $t$. The inset of Fig. \ref{fig:1}(b) shows a typical frequency spectrum of the $I(t)$ oscillations, centered on the first harmonic and measured for a dc applied voltage $V=0.4$ V at $T = 293$ K. The color map in Fig. \ref{fig:1}(b) (scale inset right) illustrates the evolution of the spectrum as $V$ changes within the NDC region. Red and green colors correspond to high signal amplitude. Note that the frequency ($f_1=1.13$ GHz) of the first harmonic depends only weakly on $V$.

\begin{figure}
\centering
 \includegraphics*[width=.8\linewidth]{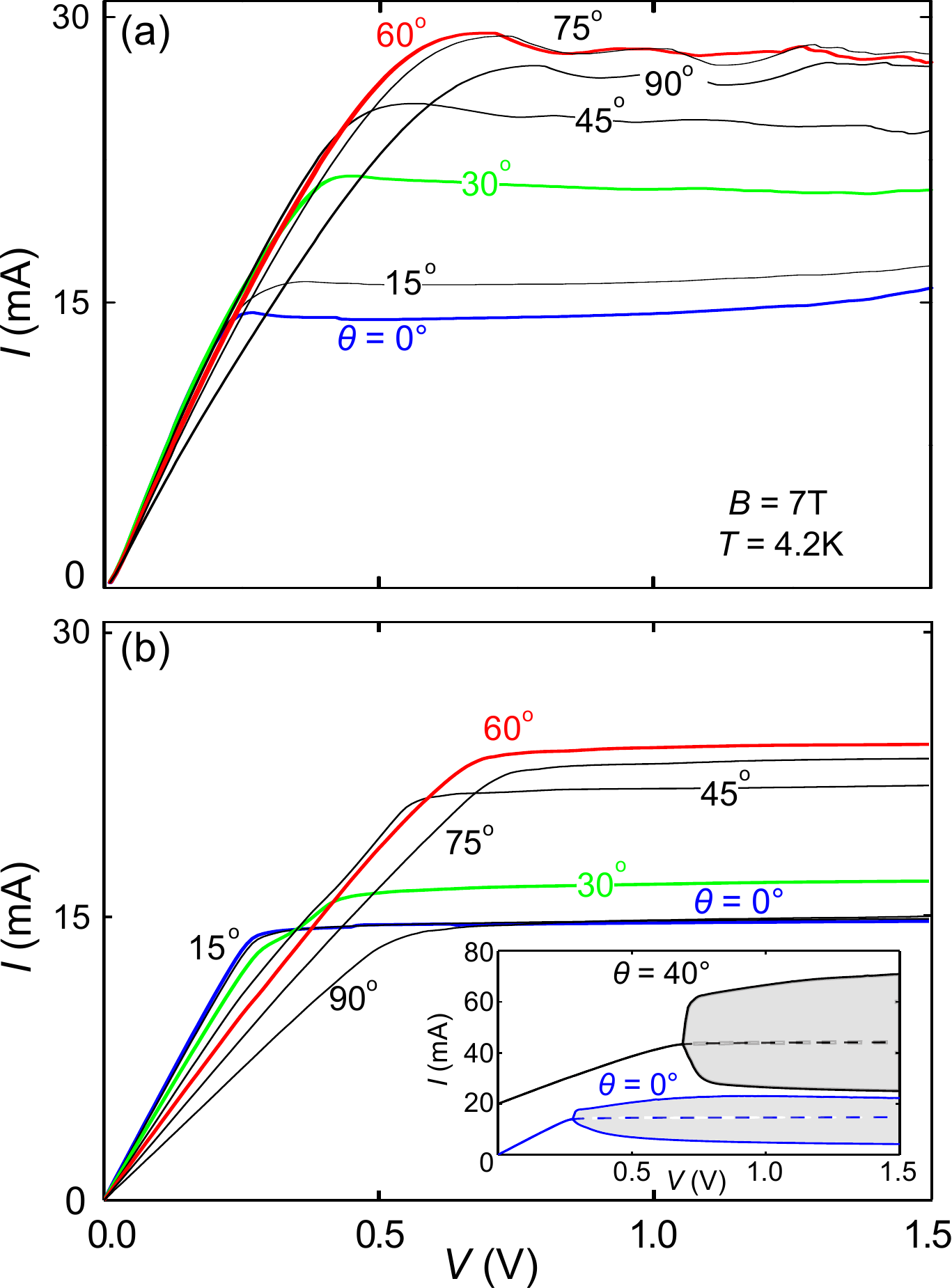}
 \caption{(Color) $I(V)$ characteristics (a) measured and (b) calculated for the SL at the $\theta$ values indicated when $B$ = 7 T and $T$ = 4.2 K. Inset in (b) shows the regions of current instability (shaded) in the $I(V)$ curves calculated for $\theta=0^{\circ}$ (lower trace) and $\theta=40^{\circ}$ (upper trace, offset vertically by 20 mA for clarity). The maximum (minimum) values of the ac current correspond to the upper (lower) edges of the shaded areas. Dashed curves in inset indicate the time-averaged current.
 \label{fig:2}}
\end{figure}

\begin{figure}
\centering
 \includegraphics*[width=.8\linewidth]{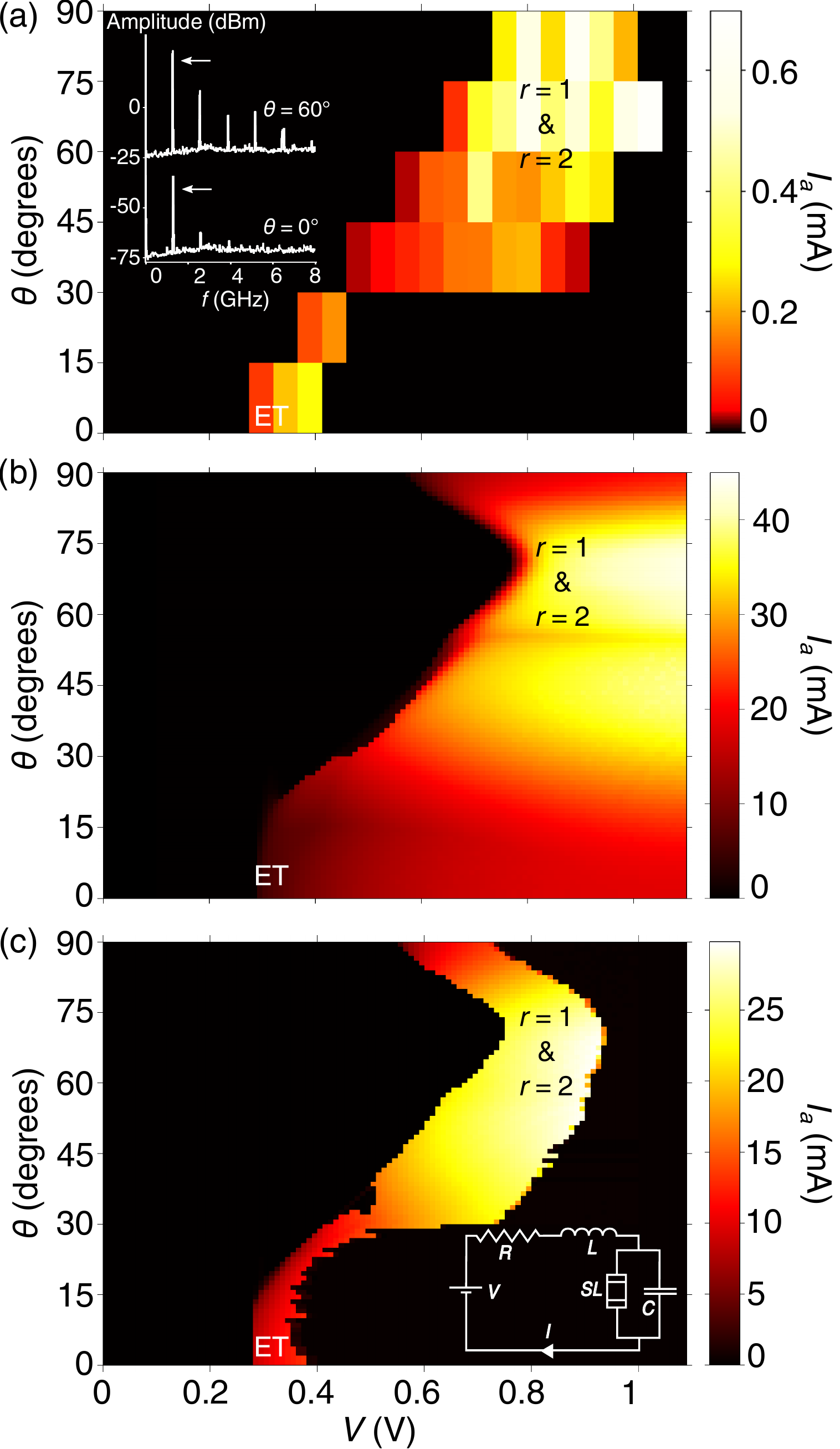}
 \caption{(Color) Color maps showing how the amplitude, $I_a$, of the $I(t)$ oscillations varies with $V$ and $\theta$ when $B$ = 7 T and $T$ = 4.2 K: (a) experimental data -- pixellation reflects discrete $V$ and $\theta$ values at which we measure $I(t)$ [inset shows amplitude of frequency spectra measured for $\theta$ = 0 at $V=0.3$ V (lower trace) and $\theta$ = $60^{\circ}$ at $V=0.8$ V (upper trace, offset vertically by 50 dBm for clarity) when $B$ = 7 T, arrows indicate first harmonics]; (b) and (c) are corresponding theoretical plots calculated, respectively, without and with the external resonant circuit shown in the inset of (c).
 \label{fig:3}}
\end{figure}

Applying a large (7 T) tilted magnetic field changes greatly both the dc and ac current. Figure \ref{fig:2}(a) shows dc $I(V)$ curves measured for a range of $\theta$ at low temperature ($T=4.2$ K). Increasing $\theta$ from $0^{\circ}$ to $60^{\circ}$ [blue and red curves in Fig. \ref{fig:2}(a)] doubles the peak current, $I_p$, from $\approx15$ mA to $\approx30$ mA. Further increasing $\theta$ to $90^{\circ}$ decreases $I_p$ to $\approx25$ mA. Figure \ref{fig:2}(a) also reveals that for $\theta \gtrsim 45^{\circ}$, the tilted magnetic field strengthens the current instability in the NDC region. This increases the amplitude, $I_a$, of the $I(t)$ oscillations, as shown in Fig. \ref{fig:3}(a). 

The inset in Fig. \ref{fig:3}(a) shows frequency spectra of the $I(t)$ oscillations measured for $\theta$ = 0 with $V$ = 0.3 V (lower trace) and $\theta$ = $60^{\circ}$ with $V$ = 0.8 V (upper trace) at $B$ = 7 T. Note that tilting the magnetic field increases the amplitude of the fundamental peaks [marked by arrows in Fig. \ref{fig:3}(a)] and higher harmonics. To explore this enhancement further, the color map in Fig. \ref{fig:3}(a) shows how the measured $I_a$ values vary with $V$ and $\theta$. The black area indicates $(V, \theta)$ values for which $I$ exhibits no temporal oscillations. The colored areas reveal two distinct islands of large-amplitude (white highest) current oscillations, labeled ``ET'' and ``$r=$ 1 \& $r=$ 2''. For $30^{\circ} \lesssim\theta \lesssim75^{\circ}$, Fig. \ref{fig:3}(a) reveals significant enhancement of $I_a$. The maximum amplitude, attained within the upper large island, is $\sim 0.7$ mA, which is larger than that measured for any $V$ when $\theta=0^{\circ}$ ($\sim 0.3$ mA). Note also that the range of $V$ over which the SL exhibits $I(t)$ oscillations [i.e. the width of the colored islands in Fig. \ref{fig:3}(a)] changes significantly with $\theta$. In particular, this range initially decreases as $\theta$ increases from $\theta=0^{\circ}$ to $\theta\approx30^{\circ}$, but thereafter increases sharply, attaining a maximum at $\theta \approx70^{\circ}$.

We model the measured effects at $T=4.2$ K by making self-consistent solutions of the drift-diffusion transport equation and Poisson's equation throughout the device -- see Eqs. (1), (2) and their detailed description in \cite{GRE2009}. For $T<50$ K, temperature changes have little effect on the electron transport \cite{selskii}. The solutions give the electron density and voltage drop versus $t$ and position $x$ throughout the SL (see Supplemental Material, \cite{Supp} Fig. S3). This model takes into account the field dependent electrical conductivity, $\sigma_c(B)=\sigma_c/[1+(\omega_c \tau_c \sin\theta)^2]$, of the contact layers, where $\sigma_c=3.8 \times 10^3$ $\Omega^{-1}$m$^{-1}$ corresponds to an electron scattering time, $\tau_c=90$ fs, in the contacts. Here, the cyclotron frequency, $\omega_c=eB/m^*$, depends on the effective electron mass, $m^*$, in GaAs. Our model also includes the electron scattering time within the SL layers, $\tau=\tau_i[\tau_e/(\tau_e+\tau_i)]^{1/2}=280$ fs determined from the elastic interface roughness scattering time $\tau_e=$ 38 fs and the inelastic scattering time $\tau_i=$ 2.1 ps. To simulate the measured dc $I(V)$ characteristics, for each value of $V$ we time average the calculated $I(t)$ curve. The resulting theoretical dc $I(V)$ curves [Fig. \ref{fig:2}(b)] are in good quantitative agreement with the experimental data [Fig. \ref{fig:2}(a)] and accurately reproduce both the voltage dependence and peak values of $I$ for all $\theta$.

\begin{figure}
\centering
 \includegraphics*[width=.8\linewidth]{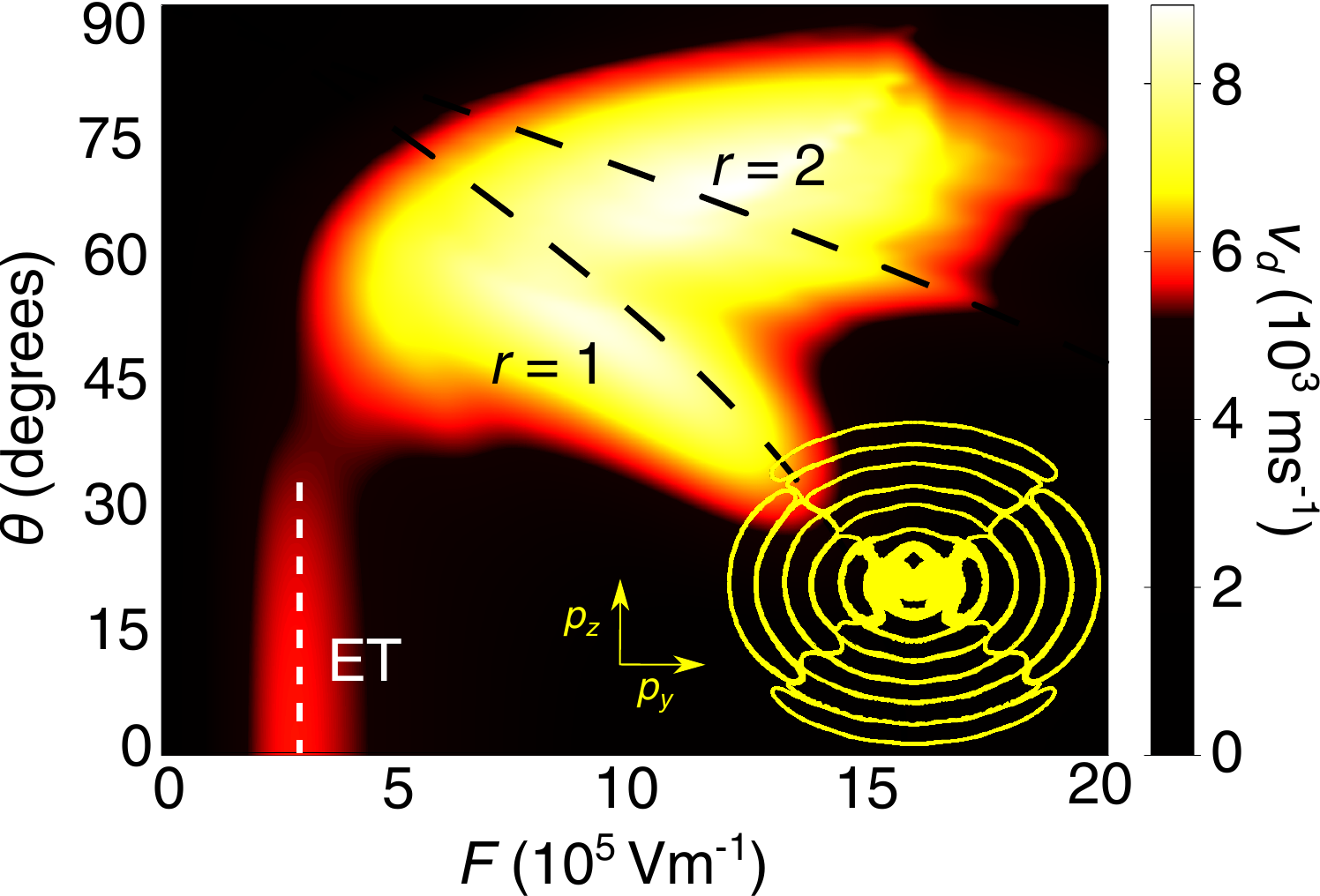}
   \caption{(Color) Color map showing the single-electron drift velocity, $v_d$, calculated versus $F$ and $\theta$ for $B$ = 7 T. Islands of high $v_d$ (whose right-hand edges are regions of negative differential velocity) originate either from Bloch oscillations (near left-hand dashed line labeled ET) or from resonant coupling of Bloch and cyclotron motion when $r$ = 1 and 2, which occurs along the middle and right-hand dashed curves respectively. The inset shows a stroboscopic Poincar\'e section of the electron trajectories constructed by plotting the crystal momentum components $(p_y,p_z)$ (yellow dots) in the plane of the SL layers at integer multiples of the Bloch oscillation period $T_B=2 \pi/ \omega_B$ when $r=2$. Resonances between the cyclotron and Bloch oscillations create unbounded web-like structures, comprising interconnecting radial and ring-shaped chaotic filaments as shown in the inset, which are known in the literature as ``stochastic webs'' \cite{Zasl91}.  
 \label{fig:4}}
\end{figure}

The inset in Fig. \ref{fig:2}(b) shows how the maximum and minimum values of the calculated $I(t)$ oscillations, marked by the upper and lower edges of the shaded windows respectively, vary with $V$ when $\theta=0^{\circ}$ (lower curve) and $40^{\circ}$ (upper curve). In each case, when $V$ is below a threshold, $V_{th}$, the maxima and minima of $I(t)$ coincide, indicating dc current. But for $V>V_{th}$, these extrema separate due to the onset of current oscillations. The time averaged current, shown for $V>V_{th}$ by the dashed curves in Fig. \ref{fig:2}(b) inset, yields the dc $I(V)$ curves shown in the main part of the figure, which agree well with the experimental data in Fig. \ref{fig:2}(a). Comparing the two traces inset in Fig. \ref{fig:2}(b) reveals that applying a tilted magnetic field significantly increases the amplitude of the ac current oscillations. This can be seen more clearly from the color map in Fig. \ref{fig:3}(b), showing the amplitude of the $I(t)$ oscillations calculated versus $V$ and $\theta$, which is similar to the experimental data in Fig. \ref{fig:3}(a). In particular, the model accurately reproduces the enhancement of the $I(t)$ oscillations observed for $30^{\circ} \lesssim\theta \lesssim75^{\circ}$ and the shapes of the islands in $(V,\theta)$ space where oscillations occur.

We now explain the physical reason for the existence and relative strength of the two islands of $I(t)$ oscillations. At critical values of $F$, $B$ and $\theta$ for which the Bloch frequency $\omega_B=eFd/\hbar=r\omega_c \cos \theta$, where $r$ is an integer, the electron orbits change abruptly from localized to unbounded chaotic paths, which propagate through an intricate mesh of conduction channels (a stochastic web \cite{Zasl91}) in phase space \cite{FROM01}: see the inset in Fig.\ref{fig:4}. This delocalization produces resonant enhancement of the electron drift velocity, $v_d$, and, hence, of $I_p$ as $\theta$ increases from $0^{\circ}$ to $60^{\circ}$. Figure \ref{fig:4} shows a color map of $v_d$ calculated versus $F$ and $\theta$ for $B$ = 7 T. The dashed curves show the loci of (left to right) the Esaki-Tsu (ET) peak, which occurs when $\omega_B =1/\tau$, and the $r=$ 1 and 2 resonances. The ET peak is visible as the lower-left light island in Fig. \ref{fig:4}. Together, the $r=$ 1 and 2 resonances generate a broader region of higher $v_d$ (upper right light area in Fig. \ref{fig:4}) for $\theta$ between $30^{\circ}$ and $75^{\circ}$. The upper and lower high $v_d$ islands in Fig. \ref{fig:4}, and the negative differential velocity (NDV) regions at their right-hand edges, trigger propagating charge domains, which generate the two corresponding islands of enhanced ac power in Fig. \ref{fig:3} (a,b).

The $I(t)$ oscillations are stronger for $30^{\circ} \lesssim \theta \lesssim 75^{\circ}$ than for $\theta=0$ for two reasons. First, $v_d$ is higher near the $r=$ 1 and 2 resonances than at the ET peak [Fig. \ref{fig:4}], which means that the charge domains move faster and, since the NDV is also stronger, contain more electrons (see Supplementary Material, \cite{Supp} Fig. S4). Second, over the voltage range corrresponding to the upper island in Fig. \ref{fig:4}, at different positions in the SL layers $F$ attains the values required for both the $r=$ 1 and 2 resonances. Consequently, propagating domains associated with both resonances coexist in the SL (see Supplementary Material, \cite{Supp} Fig. S4), thus increasing both the strength and higher-harmonic content [as shown in the upper curve inset in Fig. \ref{fig:3}(a)] of the $I(t)$ oscillations. The striking similarity between the islands of high $v_d$ in Fig. \ref{fig:4} and regions of high-amplitude current oscillations in Fig. \ref{fig:3}(a,b) clearly demonstrate that the collective electron dynamics are controlled by resonant delocalization of the underlying single-particle orbits.       

Although the above model explains how the observed amplitude of the fundamental $I(t)$ oscillation varies with $V$ and $\theta$, it overestimates the measured fundamental frequency, $f_1 = 1.13$ GHz [Figs. \ref{fig:1}(b) and \ref{fig:3}(a) inset], by an order of magnitude: specifically, it gives $f_1 \sim 10$ GHz. In addition, the model predicts that changing $V$ and/or $\theta$ can increase the fundamental frequency by a factor $\approx 10$ \cite{GRE2009}, whereas in the present experiment it is almost independent of these parameters. 

To remove this discrepancy, we modified our model to include the equivalent reactive circuit of our SL device, shown in the inset of Fig. \ref{fig:3}(c). Here, $C$, is a parasitic contact capacitance and $L$, $R$, are the inductance and resistance respectively of the equivalent circuit. 
Taking $C=2$ pF, $L=5$ nH \cite{foot1}, and $R=17$ $\Omega$, the revised model shows that, as in the experiment, the fundamental frequency of the current oscillations is $\approx$ 1 GHz for all $V$ and $\theta$. Although, for the present sample, the reactive circuit produces a ten-fold reduction in the calculated \emph{frequency} of the fundamental peak in the $I(t)$ oscillation spectrum, it has much less effect on the \emph{amplitude}, $I_a$, of the oscillations. To illustrate this, Fig. \ref{fig:3}(c) shows how $I_a$ varies with $V$ and $\theta$ when the inset equivalent circuit is incorporated in the model. The shapes and positions of the islands in this color map are in good quantitative agreement with experiment [Fig. \ref{fig:3}(a)], including the sudden enhancement of the voltage range and amplitude of the oscillations when $\theta \sim 60^{\circ}$. We attribute the smaller amplitude of the measured current oscillations to enhanced electron-electron scattering within the charge domains. This acts to reduce their electron density and, hence, the $I(t)$ oscillations.

In conclusion, we have shown that the high-frequency collective dynamics and $I(t)$ oscillations of electrons in SLs can be enhanced by using resonant delocalization of single-particle orbits to shape the $v_d(F)$ characteristics. We observe this enhancement over a broad range of $V$ and $\theta$ values for which the Bloch and cyclotron frequencies are locally commensurate within the SL. The effects of the resonant $v_d$ peaks on the collective charge-domain dynamics and resulting $I(t)$ oscillations are insensitive to external reactance. Consequently, such effects may be exploited to increase the ac power generated by the SL. They may also increase the ac frequency if the SL is designed to minimize its reactance. We emphasize that the mechanism for strengthening the current oscillations relies only on imprinting additional peaks in single-electron $v_d(F)$ curves. Since such peaks can be created by applying other signals, e.g. ac electrical \cite{HYA2009} or acoustic \cite{GRE2010} waves, the link that we have demonstrated between single and collective transport has broad implication for enhancing and controlling band transport effects for waves in other periodic systems including photonic and acoustic crystals \cite{acoustic,Ozbay}, graphene, where relativistic quantum scars of unstable periodic classical orbits were recently identified \cite{Akis}, and ultracold atoms in optical lattices \cite{Wimberger,Altman,Morsch,Mannella}.

This work is supported by the European Science Foundation, The Royal Society and EPSRC UK. We are grateful to Laurence Eaves and Kirill Alekseev for helpful discussions. We thank Mohamed Henini and Robert Airey for sample growth and fabrication.



\begin{thebibliography}{}
\bibitem{foot1}An inductance value of a few nH is expected for the gold wires (a few mm long and $\sim 100$ $\mu$m in diameter) used to contact our SL.
\end{thebibliography}
\end{document}